\documentclass[notitlepage]{article}

\usepackage{amssymb}
\usepackage{amsmath}
\usepackage[twoside,twocolumn=false]{geometry}
\usepackage[dvips]{graphicx}

\begin{document}

\title{ \textbf{Modular application of an Integration by Fractional
Expansion (IBFE) method to multiloop Feynman diagrams II}}
\author{Iv\'{a}n Gonz\'{a}lez\thanks{%
e-mail: igonzalez@fis.puc.cl} \\
Departamento de F\'{\i}sica\\
Pontificia Universidad Cat\'{o}lica de Santiago,\\
Santiago, Chile. \and Iv\'{a}n Schmidt\thanks{%
e-mail: ivan.schmidt@usm.cl} \\
Departamento de F\'{\i}sica y Centro de Estudios Subat\'{o}micos,\\
Universidad T\'{e}cnica Federico Santa Mar\'{\i}a,\\
Valparaiso, Chile}
\date{}
\maketitle

\begin{abstract}
A modular application of the integration by fractional expansion method for
evaluating Feynman diagrams is extended to diagrams that contain loop
triangle subdiagrams in their geometry. The technique is based in the
replacement of this module or subdiagram by its corresponding multiregion
expansion (MRE), which in turn is obtained from Schwinger's parametric
representation of the diagram. The result is a topological reduction,
transforming the triangular loop into an equivalent vertex, which simplifies
the search for the MRE of the complete diagram. This procedure has important
advantages with respect to considering the parametric representation of the
whole diagram: the obtained MRE is reduced and the resulting hypergeometric
series tend to have smaller multiplicity.
\end{abstract}


\bigskip


\vfill\newpage

\section{Introduction}

\qquad The integration by fractional expansion technique (IBFE) constitutes
a rather simple method that allows to evaluate Feynman diagrams starting
from the corresponding Schwinger's parametric representation, and whose main
advantage is to transfer the complexity of the direct calculation of
multidimensional integrals to the evaluation of linear systems of equations
that the method generates. This technique is particularly useful when the
propagator exponents are arbitrary, and the general solutions obtained are
sums of generalized hypergeometric functions, whose number of variables
depends on the number of invariants in the diagram and also on the
topological family to which the diagram belongs.

As is the case with any integration technique, although IBFE can be applied
to any diagram, it is not always possible to obtain simple solutions due to
the number and multivariability of the resulting hypergeometric series.
Nevertheless, for certain families of diagrams its application is
particularly simple and optimal. In a previous work \cite{IGo2} we showed
that a modular application of IBFE is a tool that simplifies considerably
the complexity of the solutions : specifically we considered bubble-type
modules or subgraphs. Here we extend this modular application to cases in
which the modules are triangular loops, which then provides a rather simple
procedure to find the solutions to a large number of Feynman diagrams.

This work is divided in the following way. In Sec. \textit{II} we apply the
IBFE technique to the massless triangular module, and find its representing
multiregion expansion for the case of off-shell external lines. Here we
define the one-loop function $G_{T}$, which allows to describe the
multiregion expansion (MRE) of this module as the sum over this loop
function and a vertex, thus effectively reducing the loop to a node. One of
the main characteristic features of the Feynman diagram general solutions,
when the propagators have arbitrary powers, is that they are given by sums
of multivariable hypergeometric functions. The arguments or variables in
these functions are ratios between invariants associated to the graph, and
given the structure of the resulting hypergeometric function, the
convergence condition requires that at least this ratio of invariants be
less than unity. Therefore from a particular diagram MRE different solutions
can be extracted, each of them associated to a particular kinematical
regime, where the differentiating element is determined by the ratio of two
invariants. This form of differentiating the solutions is direct for certain
families of diagrams \cite{IGo2}. Once all terms associated to a specific
region of interest are found, summing them will give us the desired
solution. Nevertheless, for topologies that do not belong to the previously
mentioned family, and which is the most general case, this criteria for
finding the correct terms of the solutions extracted from the MRE actually
fails, because in the general case the application of IBFE can generate
solutions which contain hypergeometric functions with a particular argument,
the unity. Given the structure of the diagram solutions, this value
corresponds to the ratio of two invariants $\left( A/B\right) $ or $\left(
B/A\right) $ , when $A=B$. Nevertheless, both ratios are associated to
different regions (different solutions) and the fact that they are equal
does not allow to differentiate them. Here we present a method that solves
this problem, and apply it to the specific case of diagrams that contain
triangle subdiagrams.

In Sec. \textit{III} we evaluate two different two-loop diagrams which
contain the triangle subgraph in their topology. The first corresponds to a
correction to a three-point function, with six propagators. We find the
general solution for arbitrary propagator powers, and then particularize it
to unity. In order to validate this technique, we compare the solution with
the one obtained from a conventional method IBP (integration by parts) \cite%
{TKa}. The second diagram is a self-energy correction with five propagators,
for which we present a new solution for the case or arbitrary values of the
propagator powers. Finally in Sec. \textit{IV} we describe some cases in
which this massless triangle function can be used and the generalization to
massive loops.

\section{Triangular module}

\qquad In this section we will find the MRE associated to the one-loop
triangular subgraph or module, considering a case in which the propagators
are massless and the external lines are off-shell. From the MRE of this
graph we will define the one-loop function $G_{T}$, which will allow to
represent this module as a particular sum of three line vertices. This
formulation will then be applied as examples to a self-energy and a
three-point function diagram, both with two loops. The self-energy example
will show quite clearly the advantages of this modular procedure.

\subsection{Obtaining the MRE of the triangular module}

We start from the momentum integral representation for this graph, which is
given in the massless case and in $D$ dimensions by:

\begin{equation}
G=\int \frac{d^{D}q}{i\pi ^{D/2}}\frac{1}{\left( (p_{1}+q)^{2}\right)
^{a_{1}}\left( (p_{1}+p_{2}+q)^{2}\right) ^{a_{2}}(q^{2})^{a_{3}}}.
\label{t7}
\end{equation}%
When the propagator powers $\left\{ a_{i}\right\} $ are integers, this
expression can be easily reduced to simpler topologies using the integration
by parts technique \cite{TKa}. The situation is more difficult for arbitrary
indices $\left\{ a_{i}\right\} $, which we are considering in this work.
Graphically the integral can be represented by the following diagram:

\begin{equation}
G=\begin{minipage}{5.9cm} \includegraphics[scale=.7] {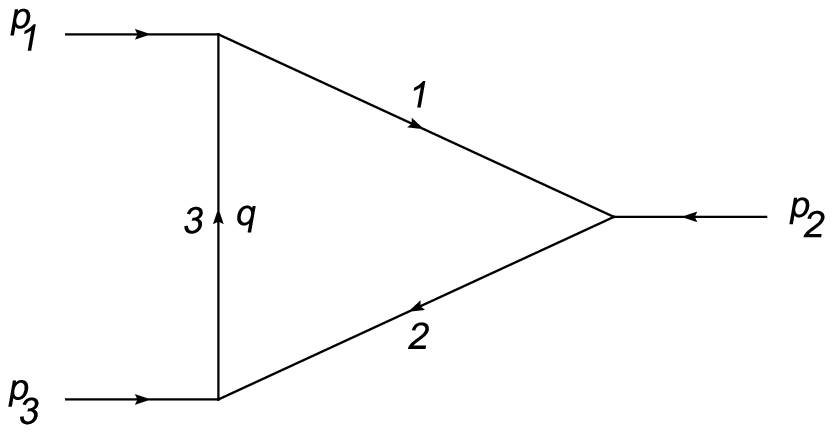} \end{minipage}
\end{equation}%
The IBFE integration technique is applied to Schwinger's parametric
representation of the diagram, which can be obtained from the matrix of
parameters \cite{IGo1}\ associated to Eq. $\left( \ref{t7}\right) $. In the
present case this matrix can be easily obtained and it is given by:

\begin{equation}
\mathbf{M}=\left(
\begin{array}{ccc}
x_{1}+x_{2}+x_{3} & x_{1}+x_{2} & x_{2} \\
x_{1}+x_{2} & x_{1}+x_{2} & x_{2} \\
x_{2} & x_{2} & x_{2}%
\end{array}%
\right) ,  \label{t4}
\end{equation}%
which allows to express Schwinger's parametric representation as:

\begin{equation}
G=\dfrac{(-1)^{-D/2}}{\Gamma (a_{1})\Gamma (a_{2})\Gamma (a_{3})}%
\int\limits_{0}^{\infty }d\overrightarrow{x}\;\frac{\exp \left( -\dfrac{%
C_{11}\;p_{1}^{2}+2C_{12}\;p_{1}.p_{2}+C_{22}\;p_{2}^{2}}{U}\right) }{U^{D/2}%
},  \label{t1}
\end{equation}%
where $d\overrightarrow{x}=x_{1}^{a_{1}-1}x_{2}^{a_{2}-1}x_{3}^{a_{3}-1}%
\;dx_{1}dx_{2}dx_{3}$ and $U=\left( x_{1}+x_{2}+x_{3}\right) $. The
coefficients $C_{ij}$ can be obtained from the determinants of the
submatrices of the matrix of parameters $\left( \ref{t4}\right) $:

\begin{equation}
\begin{array}{l}
C_{11}=\left\vert
\begin{array}{cc}
x_{1}+x_{2}+x_{3} & x_{1}+x_{2} \\
x_{1}+x_{2} & x_{1}+x_{2}%
\end{array}%
\right\vert =x_{3}(x_{1}+x_{2}), \\
\\
C_{12}=\left\vert
\begin{array}{cc}
x_{1}+x_{2}+x_{3} & x_{2} \\
x_{1}+x_{2} & x_{2}%
\end{array}%
\right\vert =x_{2}x_{3}, \\
\\
C_{22}=\left\vert
\begin{array}{cc}
x_{1}+x_{2}+x_{3} & x_{2} \\
x_{2} & x_{2}%
\end{array}%
\right\vert =x_{2}(x_{1}+x_{3}).%
\end{array}%
\end{equation}%
Replacing in $\left( \ref{t1}\right) $ and remembering that $%
p_{3}^{2}=(p_{1}+p_{2})^{2}=p_{1}^{2}+2p_{1}.p_{2}+p_{2}^{2}$, we get after
a little algebra Schwinger's parametric representation of $\left( \ref{t7}%
\right) $:

\begin{equation}
G=\dfrac{(-1)^{-D/2}}{\prod\nolimits_{j=1}^{3}\Gamma (a_{j})}%
\int\limits_{0}^{\infty }d\overrightarrow{x}\;\frac{\exp \left( -\dfrac{%
x_{3}x_{1}}{U}p_{1}^{2}\right) \exp \left( -\dfrac{x_{1}x_{2}}{U}%
p_{2}^{2}\right) \exp \left( -\dfrac{x_{2}x_{3}}{U}p_{3}^{2}\right) }{U^{D/2}%
}.  \label{t2}
\end{equation}%
Now we will deduce the MRE of the parametric integral $\left( \ref{t2}%
\right) $. For this purpose we expand the exponentials that are present in
the integrand, which leads us to the following multiple series:

\begin{equation}
G=\dfrac{(-1)^{-D/2}}{\prod\nolimits_{j=1}^{3}\Gamma (a_{j})}%
\sum\limits_{n_{1},..,n_{3}}\phi _{n_{1},..,n_{3}}\ \left( p_{1}^{2}\right)
^{n_{1}}\left( p_{2}^{2}\right) ^{n_{2}}\left( p_{3}^{2}\right)
^{n_{3}}\int\limits_{0}^{\infty }d\overrightarrow{x}\;\frac{%
x_{1}^{n_{1}+n_{2}}x_{2}^{n_{2}+n_{3}}x_{3}^{n_{1}+n_{3}}}{U^{\frac{D}{2}%
+n_{1}+n_{2}+n_{3}}},
\end{equation}%
and this allows us to obtain the MRE of the polynomial $U=\left(
x_{1}+x_{2}+x_{3}\right) $. For the denominator in the integral we have that:

\begin{equation}
\frac{1}{\left( x_{1}+x_{2}+x_{3}\right) ^{\frac{D}{2}+n_{1}+n_{2}+n_{3}}}%
=\sum\limits_{n_{4},..,n_{6}}\phi
_{n_{4},..,n_{6}}\;x_{1}^{n_{4}}x_{2}^{n_{5}}x_{3}^{n_{6}}\dfrac{%
\left\langle \frac{D}{2}+n_{1}+n_{2}+n_{3}+n_{4}+n_{5}+n_{6}\right\rangle }{%
\Gamma (\frac{D}{2}+n_{1}+n_{2}+n_{3})},
\end{equation}%
and then, integrals already separated in factors of the form $\int
dx\;x^{\alpha +...-1}$, can be replaced by the equivalent constraints $%
\left\langle \alpha +...\right\rangle $. This is the final step of the
fractional expansion process, and we have finally obtained the MRE of the
diagram from its parametric representation $\left( \ref{t2}\right) $:

\begin{equation}
\fbox{$G=\dfrac{(-1)^{-D/2}}{\prod\nolimits_{j=1}^{3}\Gamma (a_{j})}%
\sum\limits_{n_{1},..,n_{6}}\phi _{n_{1},..,n_{6}}$\ $\left(
p_{1}^{2}\right) ^{n_{1}}\left( p_{2}^{2}\right) ^{n_{2}}\left(
p_{3}^{2}\right) ^{n_{3}}\dfrac{\prod\nolimits_{j=1}^{4}\Delta _{j}}{\Gamma (%
\frac{D}{2}+n_{1}+n_{2}+n_{3})}$}  \label{t3}
\end{equation}%
Where we have summarized the constraints $\left\{ \Delta _{j}\right\} $ in
the following identities:

\begin{equation}
\left\{
\begin{array}{l}
\Delta _{1}=\left\langle \frac{D}{2}+n_{1}+n_{2}+n_{3}+n_{4}+n_{5}+n_{6}%
\right\rangle , \\
\Delta _{2}=\left\langle a_{1}+n_{1}+n_{2}+n_{4}\right\rangle , \\
\Delta _{3}=\left\langle a_{2}+n_{2}+n_{3}+n_{5}\right\rangle , \\
\Delta _{4}=\left\langle a_{3}+n_{1}+n_{3}+n_{6}\right\rangle .%
\end{array}%
\right.
\end{equation}%
From expression $\left( \ref{t3}\right) $ we see that the composition of
sums and Kronecker deltas ($\approx \Delta _{k}$) of the MRE of this module
or graph is $6\Sigma /4\delta $.

\subsection{Definition of the one-loop function $G_{T}$}

In order to generate a systematic reduction procedure of interior triangle
subgraphs of a more complex topology, we define from Eq. $\left( \ref{t3}%
\right) $ a loop function denoted $G_{T}$, such that $\left( \ref{t3}\right)
$ can be written as:

\begin{equation}
G=\sum\limits_{n_{1},..,n_{3}}G_{T}(a_{1},a_{2},a_{3};n_{1},n_{2},n_{3})\;%
\frac{1}{\left( p_{1}^{2}\right) ^{-n_{1}}}\frac{1}{\left( p_{2}^{2}\right)
^{-n_{2}}}\frac{1}{\left( p_{3}^{2}\right) ^{-n_{3}}},  \label{t8}
\end{equation}%
where the one-loop function $G_{T}(a_{1},a_{2},a_{3};n_{1},n_{2},n_{3})$ by
comparison is given by the expression:

\begin{equation}
\fbox{$G_{T}(a_{1},a_{2},a_{3};n_{1},n_{2},n_{3})=\dfrac{(-1)^{-D/2}}{%
\prod\nolimits_{j=1}^{3}\Gamma (a_{j})}\sum\limits_{n_{4},..,n_{6}}\phi
_{n_{1},..,n_{6}}$\ $\dfrac{\prod\nolimits_{j=1}^{4}\Delta _{j}}{\Gamma (%
\frac{D}{2}+n_{1}+n_{2}+n_{3})}$}
\end{equation}%
Let us look at the propagator structure of the Eq. $\left( \ref{t8}\right) $%
. It is clear that it can be represented pictorially as a sum whose argument
contains the original diagram with the loop reduced to an effective vertex
where the external lines reach, as inverse propagators. The cost of this
topological minimization is a multiple sum. Graphically the representation
of the MRE of the triangular module is given by:

\begin{equation}
\fbox{$%
\begin{minipage}{5.9cm} \includegraphics[scale=.7]
{t1.eps} \end{minipage}=\sum%
\limits_{n_{1},..,n_{3}}G_{T}(a_{1},a_{2},a_{3};n_{1},n_{2},n_{3})%
\begin{minipage}{4.6cm} \includegraphics[scale=.7]
{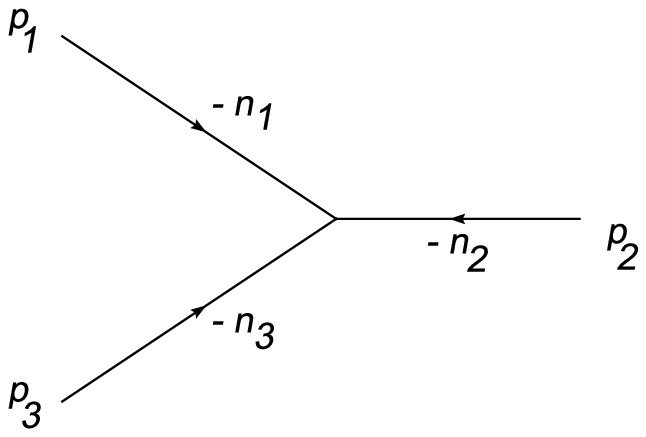} \end{minipage}$}  \label{t5}
\end{equation}%
Therefore we have shown that it is possible to reduce topologies which
contain this type of vertices, just as was done with the reduction of
propagator bubbles \cite{IGo3}.

\subsection{Method for differentiating series of unitary argument}

Before going into the actual application of IBFE to diagrams with triangle
subgraphs, it is necessary to implement a method that allows to discriminate
whether a unitary argument in the series that comes from the MRE,
corresponds to the limit of the ratio of invariants in a particular region
or to its complementary region. This is based on multiplying the external
momenta by certain constants (fictitious invariants) $A_{i}$ $\left(
i=1,2,3\right) $. Once the solution to the diagram is found, the arguments
of the obtained hypergeometric series correspond now to ratios of these
constants, allowing to separate the solutions associated to different
regions of the hypergeometric functions whose arguments are $\left( \dfrac{%
A_{i}}{A_{k}}\right) $ and $\left( \dfrac{A_{k}}{A_{i}}\right) $, with $%
i\neq k$. Once these structures have become separated, we take $%
A_{1}=A_{2}=A_{3}=1$, and the correct solution is obtained. This simple
method solves the problem of summing hypergeometric series, with the same
argument, that actually belong to different solutions of the original
problem. In the triangle case, we transform the external momenta in $\left( %
\ref{t3}\right) $ as follows:

\begin{equation}
p_{i}^{2}\Longrightarrow A_{i}p_{i}^{2}.
\end{equation}%
This transformation allows to obtain the modified expression of $\left( \ref%
{t5}\right) $:

\begin{equation}
\fbox{$%
\begin{array}{ll}
\begin{minipage}{5.9cm} \includegraphics[scale=.7]
{t1.eps} \end{minipage}= & \sum\limits_{n_{1},..,n_{3}}\left( A_{1}\right)
^{n_{1}}\left( A_{2}\right) ^{n_{2}}\left( A_{3}\right)
^{n_{3}}G_{T}(a_{1},a_{2},a_{3};n_{1},n_{2},n_{3}) \\
&  \\
& \times
\begin{minipage}{4.6cm} \includegraphics[scale=.7]
{t2.eps} \end{minipage}%
\end{array}%
$}  \label{t14}
\end{equation}

\section{Applications}

\qquad The topological reduction formula $\left( \ref{t14}\right) $, which
corresponds to an MRE of the triangular module, will be useful for finding
the MRE of generic graphs that contain in their geometry one or more of
these triangular modules. We will now explain how to use formula $\left( \ref%
{t14}\right) $ in specific cases, showing the advantages of this procedure.

\subsection{Example I : Two-loop triangle}

As the first example of the modular reduction technique and of the
topological formula $\left( \ref{t14}\right) $, let us consider a diagram $g$
with three external lines, two loops and six propagators. We take a massless
theory, with mass shell external lines $K_{1}$ and $K_{2}$ ($%
K_{1}^{2}=K_{2}^{2}=0$). The diagram is:

\begin{equation}
g=%
\begin{minipage}{6.2cm} \includegraphics[scale=.6]
{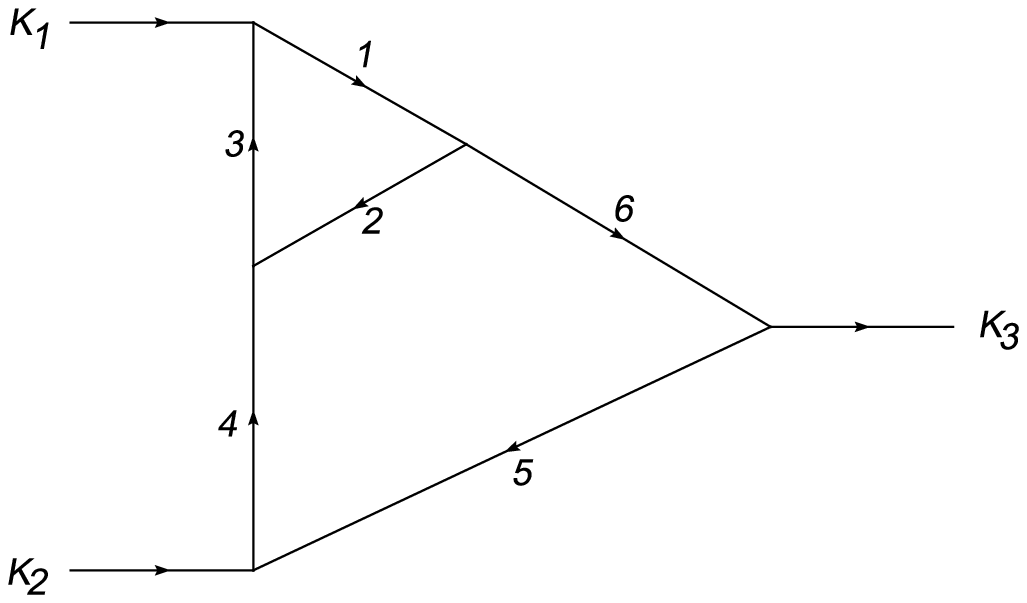} \end{minipage}  \label{t13}
\end{equation}

\subsubsection{Getting the diagram MRE}

The first step in order to find the diagram MRE with the modular procedure
is to reduce the subgraph associated to the indices $\left\{
a_{1},a_{2},a_{3}\right\} $, using formula $\left( \ref{t14}\right) $. In
this way we obtain an expression where we have eliminated one of the loops
of the diagram, which has been replaced by an effective vertex where the
three propagators associated to the indices $\left\{
-n_{1},-n_{2},-n_{3}\right\} $ meet. In detail:

\begin{equation}
\begin{array}{ll}
g= & \sum\limits_{n_{1},..,n_{3}}\left( A_{1}\right) ^{n_{1}}\left(
A_{2}\right) ^{n_{2}}\left( A_{3}\right)
^{n_{3}}G_{T}(a_{1},a_{2},a_{3};n_{1},n_{2},n_{3}) \\
&  \\
& \times
\begin{minipage}{6.2cm} \includegraphics[scale=.6]
{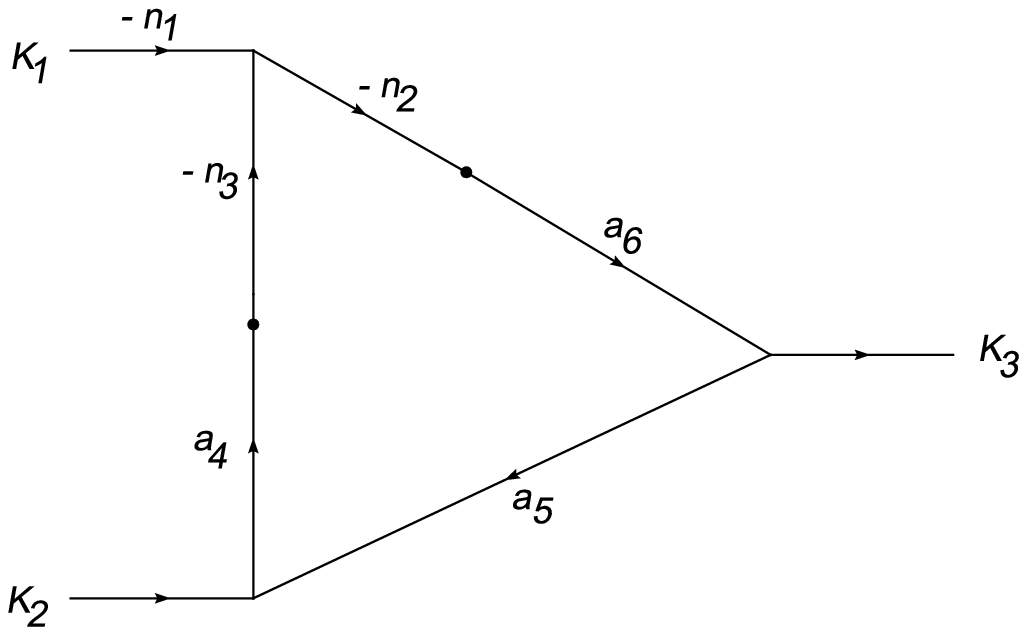} \end{minipage}%
\end{array}
\label{t10}
\end{equation}%
In Eq. $\left( \ref{t10}\right) $ we can see that one of the external lines,
the one associated with the momentum $K_{1}$, is affected by the modular
reduction on $g$. It is possible to actually extract this propagator $%
\Longrightarrow \frac{1}{\left( K_{1}^{2}\right) ^{-n_{1}}}$ from the
diagram, since it is not involved in the loop integration, and write it as a
factor that multiplies this graph. Looking at Eq. $\left( \ref{t8}\right) $,
and its pictorial equivalent $\left( \ref{t5}\right) $, we conclude that it
is possible to eliminate from the graph the information related to the index
of summation $n_{1}$. Rewriting $\left( \ref{t10}\right) $\ we thus have:

\begin{equation}
\begin{array}{ll}
g= & \sum\limits_{n_{1},..,n_{3}}\left( A_{1}\right) ^{n_{1}}\left(
A_{2}\right) ^{n_{2}}\left( A_{3}\right) ^{n_{3}}\left( K_{1}^{2}\right)
^{n_{1}}\;G_{T}(a_{1},a_{2},a_{3};n_{1},n_{2},n_{3}) \\
&  \\
& \times
\begin{minipage}{6.2cm} \includegraphics[scale=.6]
{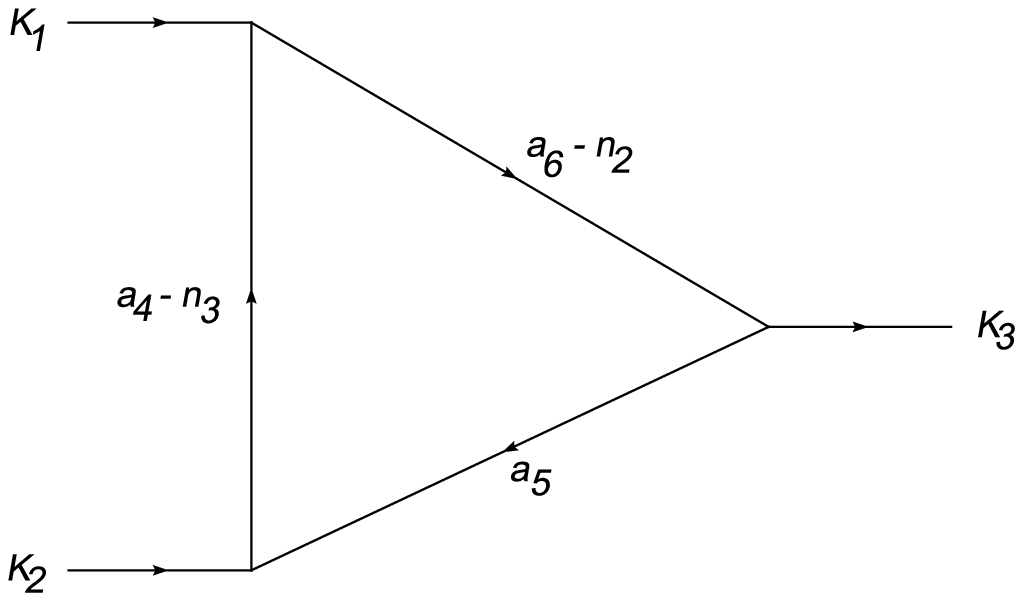} \end{minipage}%
\end{array}%
\end{equation}%
The resulting triangle allows us to apply the reduction formula $\left( \ref%
{t5}\right) $. We then get the MRE of the topology $g$ as:

\begin{equation}
\begin{array}{ll}
g= & \sum\limits_{n_{1},..,n_{3}}\left( A_{1}\right) ^{n_{1}}\left(
A_{2}\right) ^{n_{2}}\left( A_{3}\right) ^{n_{3}}\left( K_{1}^{2}\right)
^{n_{1}}\times G_{T}(a_{1},a_{2},a_{3};n_{1},n_{2},n_{3}) \\
&  \\
& \times \;G_{T}(a_{6}-n_{2},a_{5},a_{4}-n_{3};n_{7},n_{8},n_{9})\times
\begin{minipage}{4.1cm} \includegraphics[scale=.6]
{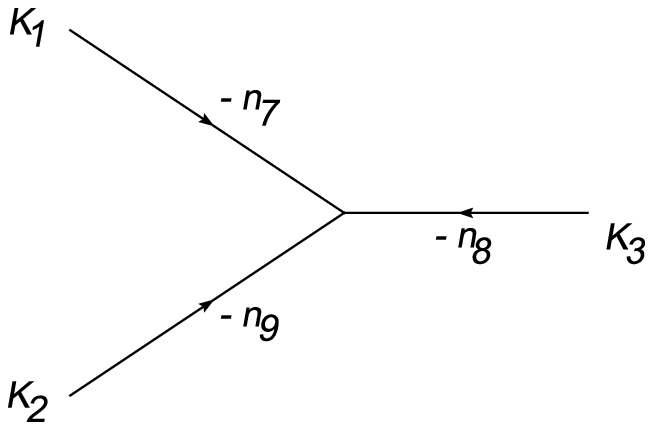} \end{minipage}%
\end{array}%
\end{equation}%
Since each function $G_{T}$ is associated to six summation indices, in order
to use the same notation for these indices $\left\{ n\right\} $, the second
function $G_{T}$ starts with the summation index $n_{7}$. Therefore we get:

\begin{equation}
\begin{array}{ll}
g= & \sum\limits_{n_{1},..,n_{3}}\left( A_{1}\right) ^{n_{1}}\left(
A_{2}\right) ^{n_{2}}\left( A_{3}\right) ^{n_{3}}\left( K_{1}^{2}\right)
^{n_{1}+n_{7}}\left( K_{2}^{2}\right) ^{n_{9}}\left( K_{3}^{2}\right)
^{n_{8}} \\
&  \\
& G_{T}(a_{1},a_{2},a_{3};n_{1},n_{2},n_{3})\times
G_{T}(a_{6}-n_{2},a_{5},a_{4}-n_{3};n_{7},n_{8},n_{9}),%
\end{array}%
\end{equation}%
where after replacing the functions $G_{T}$ by their respective MRE, we
obtain the expression:

\begin{equation}
\begin{array}{ll}
g= & \digamma \;\sum\limits_{\left\{ n\right\} }\phi _{n_{1},..,n_{12}}\
\left( A_{1}\right) ^{n_{1}}\left( A_{2}\right) ^{n_{2}}\left( A_{3}\right)
^{n_{3}}\left( K_{1}^{2}\right) ^{n_{1}+n_{7}}\left( K_{2}^{2}\right)
^{n_{9}}\left( K_{3}^{2}\right) ^{n_{8}} \\
&  \\
& \dfrac{\prod\nolimits_{j=1}^{8}\Delta _{j}}{\Gamma (\frac{D}{2}%
+n_{1}+n_{2}+n_{3})\Gamma (\frac{D}{2}+n_{7}+n_{8}+n_{9})\Gamma
(a_{6}-n_{2})\Gamma (a_{4}-n_{3})}.%
\end{array}
\label{t9}
\end{equation}%
We have defined the factor:

\begin{equation}
\digamma =\frac{\left( -1\right) ^{-D}}{\Gamma (a_{1})\Gamma (a_{2})\Gamma
(a_{3})\Gamma (a_{5})}.
\end{equation}%
The constraints $\left\{ \Delta _{j}\right\} $ are given by the following
identities:

\begin{equation}
\left\{
\begin{array}{lll}
\Delta _{1}=\left\langle \frac{D}{2}+n_{1}+n_{2}+n_{3}+n_{4}+n_{5}+n_{6}%
\right\rangle , &  & \Delta _{5}=\left\langle \frac{D}{2}%
+n_{7}+n_{8}+n_{9}+n_{10}+n_{11}+n_{12}\right\rangle , \\
\Delta _{2}=\left\langle a_{1}+n_{1}+n_{2}+n_{4}\right\rangle , &  & \Delta
_{6}=\left\langle a_{4}-n_{3}+n_{7}+n_{8}+n_{10}\right\rangle , \\
\Delta _{3}=\left\langle a_{2}+n_{2}+n_{3}+n_{5}\right\rangle , &  & \Delta
_{7}=\left\langle a_{5}+n_{8}+n_{9}+n_{11}\right\rangle , \\
\Delta _{4}=\left\langle a_{3}+n_{1}+n_{3}+n_{6}\right\rangle , &  & \Delta
_{6}=\left\langle a_{6}-n_{2}+n_{7}+n_{9}+n_{12}\right\rangle .%
\end{array}%
\right.
\end{equation}%
Finally, applying the conditions of the problem $\left(
K_{1}^{2}=K_{2}^{2}=0\right) $, the summation indices $n_{1}$, $n_{7}$ and $%
n_{9}$ must vanish in order to have a nonzero solution. Replacing these
values and eliminating the sums in $\left( \ref{t9}\right) $, we have obtain
the MRE of the diagram.

After renaming the indexes $\left\{ n\right\} \Longrightarrow \left\{
l\right\} $ so that they become consecutive, the multiregion expansion is
finally given by:

\begin{equation}
\begin{array}{ll}
g= & \digamma \;\sum\limits_{\left\{ l\right\} }\phi _{l_{1},..,l_{9}}\
\left( A_{2}\right) ^{l_{1}}\left( A_{3}\right) ^{l_{2}}\left(
K_{3}^{2}\right) ^{l_{6}} \\
&  \\
& \dfrac{\prod\nolimits_{j=1}^{8}\Delta _{j}}{\Gamma (\frac{D}{2}%
+l_{1}+l_{2})\Gamma (\frac{D}{2}+l_{6})\Gamma (a_{6}-l_{1})\Gamma
(a_{4}-l_{2})},%
\end{array}
\label{t12}
\end{equation}%
and now the constraints are given by:%
\begin{equation}
\left\{
\begin{array}{lll}
\Delta _{1}=\left\langle \frac{D}{2}+l_{1}+l_{2}+l_{3}+l_{4}+l_{5}\right%
\rangle , &  & \Delta _{5}=\left\langle \frac{D}{2}+l_{6}+l_{7}+l_{8}+l_{9}%
\right\rangle , \\
\Delta _{2}=\left\langle a_{1}+l_{1}+l_{3}\right\rangle , &  & \Delta
_{6}=\left\langle a_{4}-l_{2}+l_{6}+l_{7}\right\rangle , \\
\Delta _{3}=\left\langle a_{2}+l_{1}+l_{2}+l_{4}\right\rangle , &  & \Delta
_{7}=\left\langle a_{5}+l_{6}+l_{8}\right\rangle , \\
\Delta _{4}=\left\langle a_{3}+l_{2}+l_{5}\right\rangle , &  & \Delta
_{6}=\left\langle a_{6}-l_{1}+l_{9}\right\rangle .%
\end{array}%
\right.
\end{equation}%
Therefore the multiplicity of the hypergeometric series that can be
extracted from the MRE is one $\Longrightarrow $ hypergeometric series of
the type \ $_{q}F_{\left( q-1\right) }$ and at most $C_{8}^{9}=9$ series of
this type whose argument for this case is one.

\subsubsection{General IBFE solution}

Now we proceed to find the general solutions associated with the diagram $%
\left( \ref{t13}\right) $ coming from the MRE $\left( \ref{t12}\right) $\
which represents it. We maintain all the indexes $\left\{ a_{j}\right\} $\
with arbitrary values, and for notational simplicity make $A_{2}=A$ and $%
A_{3}=B$.

The solutions we present here correspond to two regions that can be
differentiated according to the ratio of the fictitious invariants used in
order to separate them: the region where $A>B$ and the region where $B>A$.
Both are solutions of $g$, related by analytical continuation. In
particular, in the limit $A=B=1$ they become identical, as will be seen
later on.

\paragraph{Analytical solution in the region $\left\vert \dfrac{A}{B}%
\right\vert \leq 1$}

Let us first define the following simplifying notation for the indexes $%
\left\{ a_{j}\right\} $:

\begin{equation}
a_{ijk...}=a_{i}+a_{j}+a_{k}+....
\end{equation}%
Moreover, since in practice each constraint $\left\langle ...\right\rangle $%
\ eliminates a sum in the MRE, the remaining or free sum at the end of the
elimination process generates the corresponding hypergeometric
representation $\;_{q}F_{q-1}$\thinspace\ which we will identify as $G_{j}$,
since it is the contribution obtained when the summation index $n_{j}$ is
free in the multiregion representation of diagram $g$.

With these definitions we can write the solutions for this region as:

\begin{equation}
g=g\left( \tfrac{A}{B}\right) =G_{1}+G_{5},
\end{equation}%
where the terms inside the sum correspond to the following functions:

\begin{equation}
\begin{array}{ll}
G_{1}= & (-1)^{-D}\left( K_{3}^{2}\right) ^{D-a_{123456}}\dfrac{\Gamma
(a_{123}-\frac{D}{2})\Gamma (a_{123456}-D)\Gamma (\frac{D}{2}-a_{13})\Gamma (%
\frac{D}{2}-a_{12})\Gamma (\frac{D}{2}-a_{56})\Gamma (D-a_{12346})}{\Gamma
(a_{2})\Gamma (a_{3})\Gamma (a_{5})\Gamma (D-a_{123})\Gamma (a_{1234}-\frac{D%
}{2})\Gamma (\frac{3D}{2}-a_{123456})} \\
&  \\
& \times B^{\frac{D}{2}-a_{123}}\;_{3}F_{2}\left( \left.
\begin{array}{c}
\begin{array}{ccc}
a_{123}-\tfrac{D}{2}, & a_{1}, & \frac{D}{2}-a_{56}%
\end{array}
\\
\begin{array}{cc}
a_{1234}-\frac{D}{2}, & 1+a_{12}-\frac{D}{2}%
\end{array}%
\end{array}%
\right\vert \dfrac{A}{B}\right) ,%
\end{array}%
\end{equation}%
and:

\begin{equation}
\begin{array}{ll}
G_{5}= & (-1)^{-D}\left( K_{3}^{2}\right) ^{D-a_{123456}}\dfrac{\Gamma
(a_{12}-\frac{D}{2})\Gamma (\frac{D}{2}-a_{2})\Gamma (\frac{D}{2}%
-a_{13})\Gamma (D-a_{12346})\Gamma (a_{123456}-D)\Gamma (D-a_{1256})}{\Gamma
(a_{1})\Gamma (a_{2})\Gamma (a_{5})\Gamma (D-a_{123})\Gamma (a_{34})\Gamma (%
\frac{3D}{2}-a_{123456})} \\
&  \\
& \times A^{\frac{D}{2}-a_{12}}B^{-a_{3}}\;_{3}F_{2}\left( \left.
\begin{array}{c}
\begin{array}{ccc}
D-a_{1256}, & a_{3}, & \frac{D}{2}-a_{2}%
\end{array}
\\
\begin{array}{cc}
a_{34}, & \frac{D}{2}+1-a_{12}%
\end{array}%
\end{array}%
\right\vert \dfrac{A}{B}\right) .%
\end{array}%
\end{equation}

\paragraph{Analytical solution in the region $\left\vert \dfrac{B}{A}%
\right\vert \leq 1$}

Analogously we have that in this fictitious region the solution is given by:

\begin{equation}
g=g\left( \tfrac{B}{A}\right) =G_{2}+G_{3}+G_{7},
\end{equation}%
with:

\begin{equation}
\begin{array}{ll}
G_{2}= & (-1)^{-D}\left( K_{3}^{2}\right) ^{D-a_{123456}}\dfrac{\Gamma
(a_{123}-\frac{D}{2})\Gamma (\frac{D}{2}-a_{23})\Gamma (\frac{D}{2}%
-a_{13})\Gamma (a_{123456}-D)\Gamma (D-a_{12356})\Gamma (D-a_{12346})}{%
\Gamma (a_{1})\Gamma (a_{2})\Gamma (a_{5})\Gamma (a_{4})\Gamma
(D-a_{123})\Gamma (\frac{3D}{2}-a_{123456})} \\
&  \\
& \times A^{\frac{D}{2}-a_{123}}\;_{3}F_{2}\left( \left.
\begin{array}{c}
\begin{array}{ccc}
a_{123}-\frac{D}{2}, & 1-a_{4}, & a_{3}%
\end{array}
\\
\begin{array}{cc}
1-\frac{D}{2}+a_{23}, & 1-\frac{D}{2}+a_{12356}%
\end{array}%
\end{array}%
\right\vert \dfrac{B}{A}\right) ,%
\end{array}%
\end{equation}

\bigskip

\begin{equation}
\begin{array}{ll}
G_{3}= & (-1)^{-D}\left( K_{3}^{2}\right) ^{D-a_{123456}}\dfrac{\Gamma
(a_{23}-\frac{D}{2})\Gamma (\frac{D}{2}-a_{13})\Gamma (\frac{D}{2}%
-a_{2})\Gamma (a_{123456}-D)\Gamma (\frac{D}{2}-a_{156})\Gamma (D-a_{12346})%
}{\Gamma (a_{2})\Gamma (a_{3})\Gamma (a_{5})\Gamma (D-a_{123})\Gamma
(a_{234}-D)\Gamma (\frac{3D}{2}-a_{123456})} \\
&  \\
& \times A^{-a_{1}}B^{\frac{D}{2}-a_{23}}\;_{3}F_{2}\left( \left.
\begin{array}{c}
\begin{array}{ccc}
a_{1}, & \frac{D}{2}+1-a_{24}, & \frac{D}{2}-a_{2}%
\end{array}
\\
\begin{array}{cc}
\frac{D}{2}+1-a_{23}, & 1+a_{156}-\frac{D}{2}%
\end{array}%
\end{array}%
\right\vert \dfrac{B}{A}\right) ,%
\end{array}%
\end{equation}%
and finally:

\begin{equation}
\begin{array}{ll}
G_{7}= & (-1)^{-D}\left( K_{3}^{2}\right) ^{D-a_{123456}}\dfrac{\Gamma
(a_{23}-\frac{D}{2})\Gamma (\frac{D}{2}-a_{13})\Gamma (\frac{D}{2}%
-a_{2})\Gamma (a_{123456}-D)\Gamma (\frac{D}{2}-a_{156})\Gamma (D-a_{12346})%
}{\Gamma (a_{2})\Gamma (a_{3})\Gamma (a_{5})\Gamma (D-a_{123})\Gamma
(a_{234}-D)\Gamma (\frac{3D}{2}-a_{123456})} \\
&  \\
& \times A^{a_{56}-\frac{D}{2}}B^{D-a_{12356}}\;_{3}F_{2}\left( \left.
\begin{array}{c}
\begin{array}{ccc}
D+1-a_{123456}, & \frac{D}{2}-a_{56}, & D-a_{1256}%
\end{array}
\\
\begin{array}{cc}
D+1-a_{12356}, & \frac{D}{2}+1-a_{156}%
\end{array}%
\end{array}%
\right\vert \dfrac{B}{A}\right) .%
\end{array}%
\end{equation}

\subsubsection{Particular case : unit indexes}

In the actual Feynman diagram evaluation, in general the corresponding
momentum integral has indexes or propagator powers that are equal to unity.
From the previous results, making $a_{j}=1\left( j=1,...,6\right) $, we
obtain the solution:

\paragraph{Analytic solution in the region $\left\vert \dfrac{A}{B}%
\right\vert \leq 1$}

The solutions in this region are given by:

\begin{equation}
g=g\left( \tfrac{A}{B}\right) =G_{1}+G_{5},  \label{t15}
\end{equation}%
where now:

\begin{equation}
G_{1}=(-1)^{-D}\left( K_{3}^{2}\right) ^{D-6}\dfrac{\Gamma (3-\frac{D}{2}%
)\Gamma (6-D)\Gamma (\frac{D}{2}-2)^{3}\Gamma (D-5)}{\Gamma (D-3)\Gamma (4-%
\frac{D}{2})\Gamma (\frac{3D}{2}-6)}\times B^{\frac{D}{2}-3}\;_{2}F_{1}%
\left( \left.
\begin{array}{c}
\begin{array}{cc}
1, & \frac{D}{2}-2%
\end{array}
\\
4-\frac{D}{2}%
\end{array}%
\right\vert \dfrac{A}{B}\right) ,
\end{equation}%
and:

\begin{equation}
\begin{array}{ll}
G_{5}= & (-1)^{-D}\left( K_{3}^{2}\right) ^{D-6}\dfrac{\Gamma (2-\frac{D}{2}%
)\Gamma (\frac{D}{2}-1)\Gamma (\frac{D}{2}-2)\Gamma (D-5)\Gamma (6-D)\Gamma
(D-4)}{\Gamma (D-3)\Gamma (\frac{3D}{2}-6)} \\
&  \\
& \times A^{\frac{D}{2}-2}B^{-1}\;_{2}F_{1}\left( \left.
\begin{array}{c}
\begin{array}{cc}
D-4, & 1%
\end{array}
\\
2%
\end{array}%
\right\vert \dfrac{A}{B}\right) .%
\end{array}%
\end{equation}

\paragraph{Analytical solution in the region $\left\vert \dfrac{B}{A}%
\right\vert \leq 1$}

In this region the solution is:

\begin{equation}
g=g\left( \tfrac{B}{A}\right) =G_{2}+G_{3}+G_{7},  \label{t16}
\end{equation}%
where the hypergeometric representations $G_{2,3,7}$\ are now given by:

\begin{equation}
G_{2}=(-1)^{-D}\left( K_{3}^{2}\right) ^{D-6}\dfrac{\Gamma (3-\frac{D}{2}%
)\Gamma (\frac{D}{2}-2)^{2}\Gamma (6-D)\Gamma (D-5)^{2}}{\Gamma (D-3)\Gamma (%
\frac{3D}{2}-6)}\times A^{\frac{D}{2}-3}\;_{2}F_{1}\left( \left.
\begin{array}{c}
\begin{array}{cc}
0, & 1%
\end{array}
\\
6-D%
\end{array}%
\right\vert \dfrac{B}{A}\right) ,
\end{equation}%
or equivalently:

\begin{equation}
G_{2}=(-1)^{-D}\left( K_{3}^{2}\right) ^{D-6}\dfrac{\Gamma (3-\frac{D}{2}%
)\Gamma (\frac{D}{2}-2)^{2}\Gamma (6-D)\Gamma (D-5)^{2}}{\Gamma (D-3)\Gamma (%
\frac{3D}{2}-6)}\times A^{\frac{D}{2}-3},
\end{equation}

\bigskip

\begin{equation}
\begin{array}{ll}
G_{3}= & (-1)^{-D}\left( K_{3}^{2}\right) ^{D-6}\dfrac{\Gamma (2-\frac{D}{2}%
)\Gamma (\frac{D}{2}-2)\Gamma (\frac{D}{2}-1)\Gamma (6-D)\Gamma (D-5)\Gamma (%
\frac{D}{2}-3)}{\Gamma (D-3)\Gamma (\frac{3D}{2}-6)\Gamma (3-\frac{D}{2})}
\\
&  \\
& \times A^{-1}B^{\frac{D}{2}-2}\;_{2}F_{1}\left( \left.
\begin{array}{c}
\begin{array}{cc}
1, & \frac{D}{2}-2%
\end{array}
\\
4-\frac{D}{2}%
\end{array}%
\right\vert \dfrac{B}{A}\right) ,%
\end{array}%
\end{equation}%
and finally:

\begin{equation}
\begin{array}{ll}
G_{7}= & (-1)^{-D}\left( K_{3}^{2}\right) ^{D-6}\dfrac{\Gamma (2-\frac{D}{2}%
)\Gamma (\frac{D}{2}-2)\Gamma (\frac{D}{2}-1)\Gamma (6-D)\Gamma (\frac{D}{2}%
-3)\Gamma (D-5)}{\Gamma (D-3)\Gamma (\frac{3D}{2}-6)} \\
&  \\
& \times A^{2-\frac{D}{2}}B^{D-5}\;_{1}F_{0}\left( \left.
\begin{array}{c}
D-5 \\
-%
\end{array}%
\right\vert \dfrac{B}{A}\right) .%
\end{array}%
\end{equation}

\subsubsection{Analytic continuation and dependence between the solutions
obtained with IBFE}

It is possible to show that in the limit $A=B=1$ the solutions $\left( \ref%
{t15}\right) $\ and $\left( \ref{t16}\right) $\ are equal, and therefore
both are related by analytic continuation. For this purpose we use the
following identity:

\begin{equation}
\;_{2}F_{1}\left( \left.
\begin{array}{c}
\begin{array}{cc}
a, & b%
\end{array}
\\
c%
\end{array}%
\right\vert 1\right) =\dfrac{\Gamma (c)\Gamma (c-a-b)}{\Gamma (c-a)\Gamma
(c-b)}.  \label{t17}
\end{equation}%
Using this formula we easily find that the summed series have the following
form:

\begin{equation}
G_{1}=(-1)^{-D}\left( K_{3}^{2}\right) ^{D-6}\dfrac{\Gamma (\frac{D}{2}%
-2)^{3}\Gamma (D-5)\Gamma (5-D)}{\Gamma (D-3)\Gamma (\frac{3D}{2}-6)},
\end{equation}%
y:

\begin{equation}
G_{5}=(-1)^{-D}\left( K_{3}^{2}\right) ^{D-6}\dfrac{\Gamma (2-\frac{D}{2}%
)\Gamma (\frac{D}{2}-2)\Gamma (\frac{D}{2}-1)\Gamma (D-4)\Gamma (D-5)\Gamma
(5-D)}{\Gamma (D-3)\Gamma (\frac{3D}{2}-6)}.
\end{equation}%
On the other hand we have that:

\begin{equation}
G_{2}=(-1)^{-D}\left( K_{3}^{2}\right) ^{D-6}\dfrac{\Gamma (3-\frac{D}{2}%
)\Gamma (\frac{D}{2}-2)^{2}\Gamma (6-D)\Gamma (D-5)^{2}}{\Gamma (D-3)\Gamma (%
\frac{3D}{2}-6)},
\end{equation}%
and :

\begin{equation}
G_{3}=(-1)^{-D}\left( K_{3}^{2}\right) ^{D-6}\dfrac{\Gamma (2-\frac{D}{2}%
)\Gamma (\frac{D}{2}-2)\Gamma (\frac{D}{2}-1)\Gamma (D-5)\Gamma (\frac{D}{2}%
-3)\Gamma \left( 4-\frac{D}{2}\right) \Gamma \left( 5-D\right) }{\Gamma
(D-3)\Gamma (\frac{3D}{2}-6)\Gamma (3-\frac{D}{2})^{2}}.
\end{equation}%
The hypergeometric function contained in the term $G_{7}$ and which has the
form $\;_{1}F_{0}$, can be transformed in another one of the type $%
\;_{2}F_{1}$, and then we can apply the identity $\left( \ref{t17}\right) $,

\begin{equation}
\;_{1}F_{0}\left( \left.
\begin{array}{c}
\alpha \\
-%
\end{array}%
\right\vert 1\right) =\;_{2}F_{1}\left( \left.
\begin{array}{c}
\begin{array}{cc}
\alpha , & \beta%
\end{array}
\\
\beta%
\end{array}%
\right\vert 1\right) ,
\end{equation}%
and then we get that

\begin{equation}
G_{7}=0.
\end{equation}%
After a bit of algebra, and using known Gamma function properties, it is
straightforward to show that the solutions $\left( \ref{t15}\right) $ and $%
\left( \ref{t16}\right) $ are equal in the limit $A=B=1$, giving

\begin{equation}
\left( G_{1}+G_{5}\right) -\left( G_{2}+G_{3}\right) =0.
\end{equation}%
Therefore, defining $G_{IBFE}$ as the solution for the graph in $\left( \ref%
{t13}\right) $, the final solution in general can be written as:

\begin{equation}
G_{IBFE}=\left\{
\begin{array}{l}
G_{2}+G_{3} \\
\\
\text{or} \\
\\
G_{1}+G_{5}.%
\end{array}%
\right.  \label{t18}
\end{equation}

\subsubsection{Validating the solution $G_{IBFE}$. Comparison with a
conventional method : Integration by parts (IBP)}

When the propagator powers are integers, it is possible to evaluate such
topologies with the integration by parts method \cite{TKa}. In the
particular case of diagrams that contain subgraphs, this technique can be
quite useful in order to simplify the integration problem, since it allows
to simplify the geometry of the diagram.

In the particular case of the diagram which contain triangle subgraphs, this
technique can be in order to simplify the integration problem. One very
useful formula, which can be deduced applying IBP to a triangle graph, and
which is known as the triangle identity, is the following:

\begin{equation}
\begin{array}{rr}
\begin{minipage}{2.9cm} \includegraphics[scale=.6]
{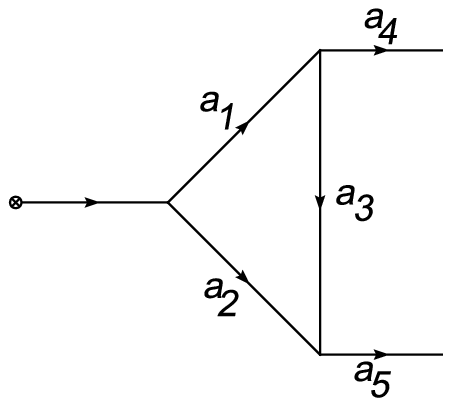} \end{minipage}= & \dfrac{1}{\left(
D-a_{1}-a_{2}-2a_{3}\right) }\left[ \;a_{1}\left(
\begin{minipage}{3.0cm} \includegraphics[scale=.6]
{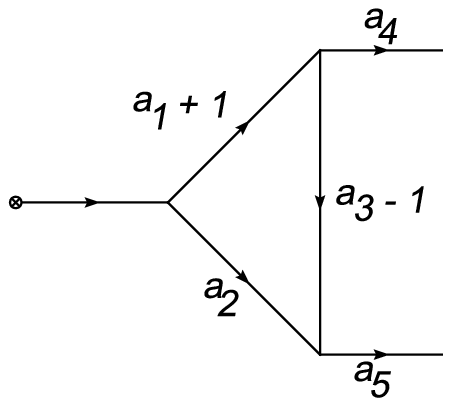} \end{minipage}-%
\begin{minipage}{3.0cm} \includegraphics[scale=.6]
{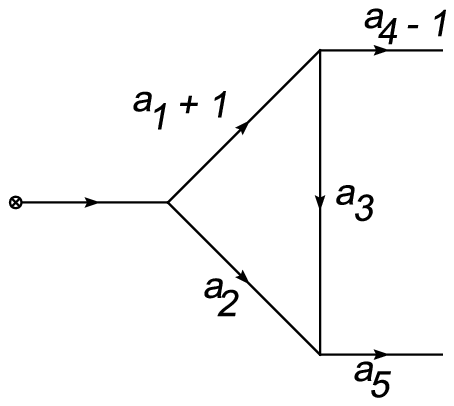} \end{minipage}\;\right) \right. \\
&  \\
& \left. +\;a_{2}\left(
\begin{minipage}{3.0cm} \includegraphics[scale=.6]
{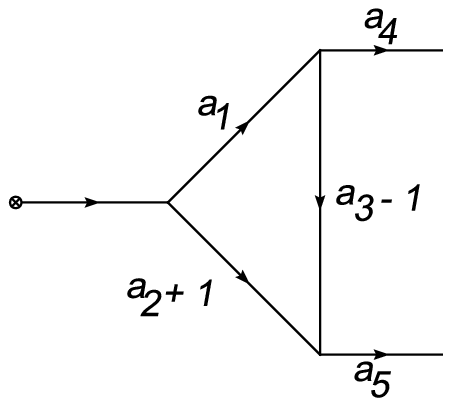} \end{minipage}-%
\begin{minipage}{3.0cm} \includegraphics[scale=.6]
{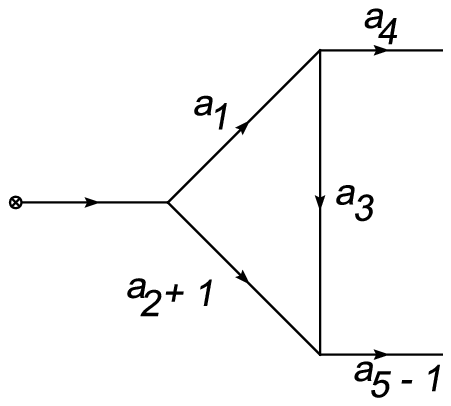} \end{minipage}\;\right) \;\right] .%
\end{array}
\label{Rt}
\end{equation}%
where it has been assumed that the indexes $a_{1},a_{2},$ and $a_{3}$ have
arbitrary values. We now apply directly this identity to the diagram
depicted in $\left( \ref{t13}\right) $, but taking all the propagator powers
to be unity. The result that we obtain is:

\begin{equation}
\begin{array}{ll}
g & =%
\begin{minipage}{3.0cm} \includegraphics[scale=.6]
{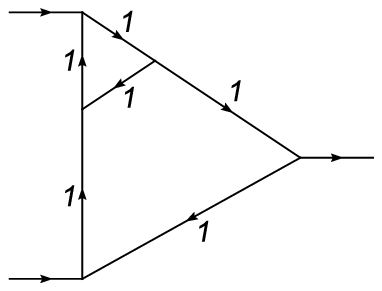} \end{minipage} \\
&  \\
& =\dfrac{1}{D-4}\left[ 2\left(
\begin{minipage}{3.0cm} \includegraphics[scale=.6]
{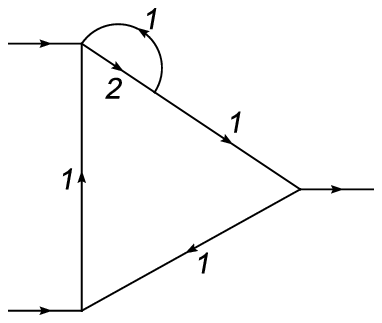} \end{minipage}\right) -%
\begin{minipage}{3.0cm} \includegraphics[scale=.6]
{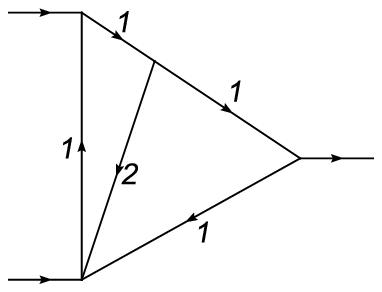} \end{minipage}-%
\begin{minipage}{3.0cm} \includegraphics[scale=.6]
{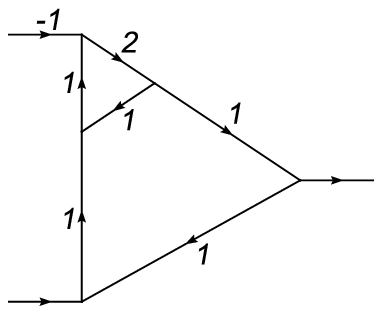} \end{minipage}\right] .%
\end{array}%
\end{equation}%
The third diagram in the left-hand side of the equation vanishes (external
on-shell line condition). The second diagram can be further simplified using
again $\left( \ref{Rt}\right) $, giving:

\begin{equation}
\begin{minipage}{2.6cm} \includegraphics[scale=.65]
{tibp3.eps} \end{minipage}=\frac{1}{D-5}\left[
\begin{minipage}{2.8cm} \includegraphics[scale=.65]
{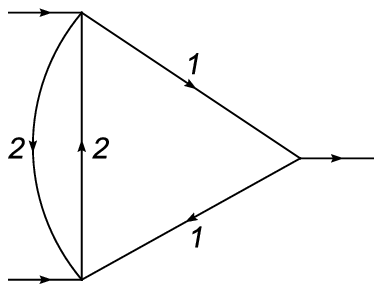} \end{minipage}+2\left(
\begin{minipage}{3.0cm} \includegraphics[scale=.6]
{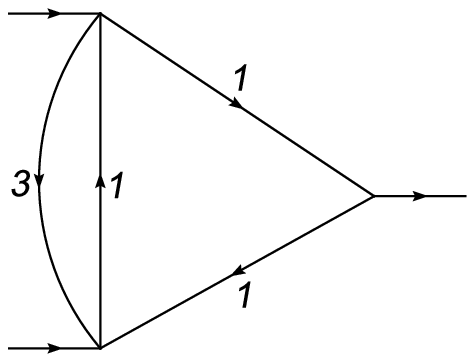} \end{minipage}-%
\begin{minipage}{3.0cm} \includegraphics[scale=.6]
{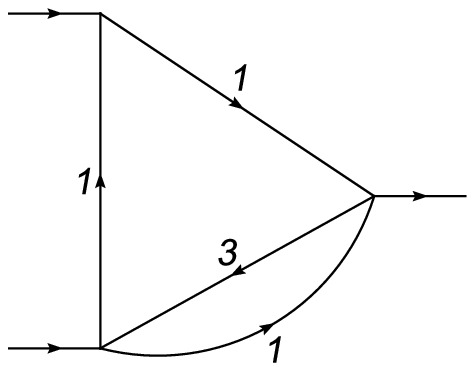} \end{minipage}\right) \right] .
\end{equation}%
Now each term of the solution can be easily rewritten evaluating the bubble
diagram and then the triangle. For this we need the general formulae for
these graphs, given by:

\begin{equation}
\begin{array}{cc}
a_{1} &  \\
\begin{minipage}{4.1cm} \includegraphics[scale=.6]
{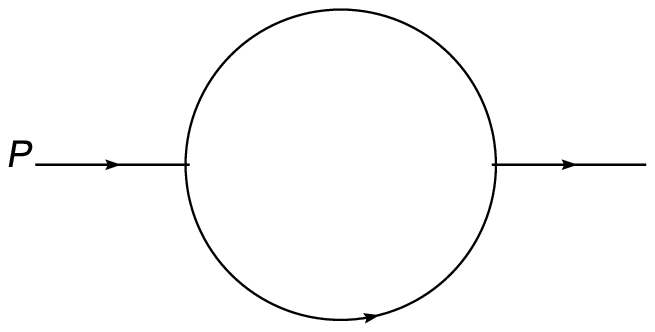} \end{minipage} & =\;\mathbf{G}(a_{1},a_{2})\;\dfrac{1}{\left(
P^{2}\right) ^{a_{1}+a_{2}-\frac{D}{2}}} \\
a_{2} &
\end{array}
\label{t19}
\end{equation}%
where:

\begin{equation}
\mathbf{G}(a_{1},a_{2})=(-1)^{-D/2}\dfrac{\Gamma (a_{1}+a_{2}-\frac{D}{2}%
)\Gamma (\frac{D}{2}-a_{1})\Gamma (\frac{D}{2}-a_{2})}{\Gamma (a_{1})\Gamma
(a_{2})\Gamma (D-a_{1}-a_{2})}.
\end{equation}%
In the case of the triangle with conditions $P_{1}^{2}=P_{2}^{2}=0$ one has
the fundamental formula:

\begin{equation}
\begin{minipage}{5.6cm} \includegraphics[scale=.6]
{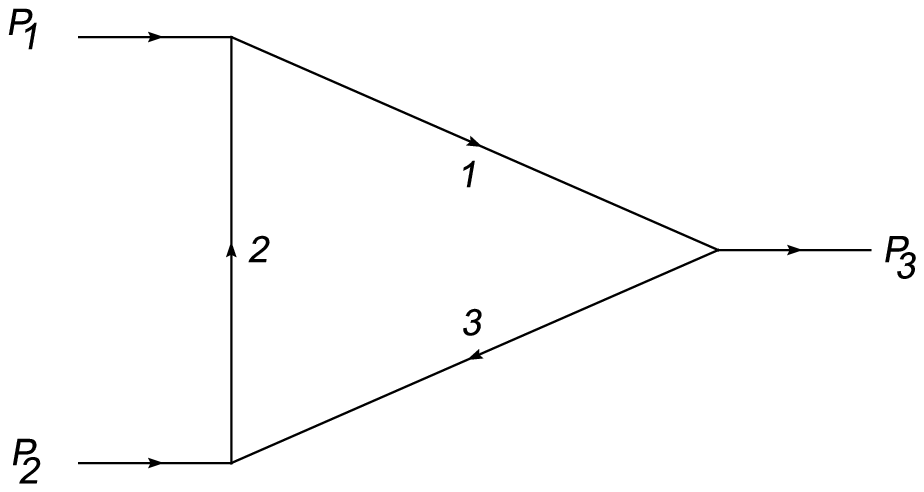} \end{minipage}=\mathbf{G}(a_{1},a_{2},a_{3})\;\left(
P_{3}^{2}\right) ^{\frac{D}{2}-a_{1}-a_{3}-a_{2}}  \label{t20}
\end{equation}%
where the factor $\mathbf{G}(a_{1},a_{2},a_{3})$ is given by:

\begin{equation}
\mathbf{G}(a_{1},a_{2},a_{3})=(-1)^{-D/2}\dfrac{\Gamma (a_{1}+a_{2}+a_{3}-%
\frac{D}{2})\Gamma (\frac{D}{2}-a_{2}-a_{3})\Gamma (\frac{D}{2}-a_{1}-a_{2})%
}{\Gamma (a_{1})\Gamma (a_{3})\Gamma (D-a_{1}-a_{3}-a_{2})}
\end{equation}%
Applying formulae $\left( \ref{t19}\right) $\ and $\left( \ref{t20}\right) $%
, the solution for the diagram is then:

\begin{equation}
\begin{array}{ll}
g=G_{IBP}= & \dfrac{\left( K_{3}^{2}\right) ^{D-6}}{D-4}\left[ 2\mathbf{G}%
(1,2)\mathbf{G}(4-\tfrac{D}{2},1,1)-\dfrac{1}{D-5}\mathbf{G}(2,2)\mathbf{G}%
(1,4-\tfrac{D}{2},1)\right. \\
&  \\
& \left. -\dfrac{1}{D-5}\;2\mathbf{G}(1,3)\left[ \mathbf{G}(1,4-\tfrac{D}{2}%
,1)-\mathbf{G}(1,1,4-\tfrac{D}{2})\right] \right]%
\end{array}
\label{sol_IBP}
\end{equation}%
Choosing one of the solutions for $G_{IBFE}$ $\left( \ref{t18}\right) $, it
can be easily shown that the solution of this diagram evaluated using the
IBFE technique gives the same result as with IBP:%
\begin{equation}
G_{IBFE}=G_{IBP}.
\end{equation}%
This is then a concrete confirmation of the IBFE technique, applied
modularly to a diagram with triangle subgraphs. The important point is that
this has allowed to obtain the MRE of the graph in a noticeably more direct
and systematic way than finding it by taking Schwinger%
\'{}%
s parametric representation of the whole diagram.

\subsection{Example II : Two-loop propagator correction}

A more relevant example is the two-loop propagator correction, with five
internal lines.%
\begin{equation}
G=%
\begin{minipage}{4.5cm} \includegraphics[scale=.7]
{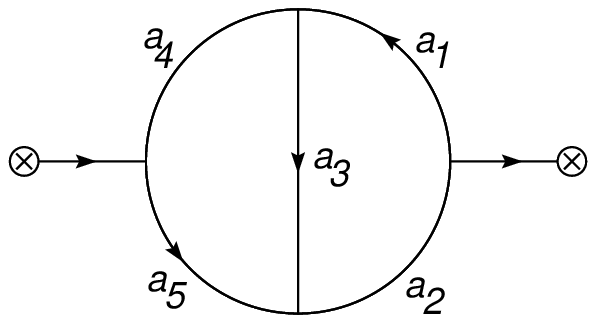} \end{minipage}
\end{equation}

\subsubsection{Obtaining the diagram MRE}

In order to find the diagram MRE we apply formula $\left( \ref{t14}\right) $
on the left-hand side of the diagram, obtaining the following graphical
expression:

\begin{equation}
G=\sum\limits_{n_{1},..,n_{3}}\left( A_{1}\right) ^{n_{1}}\left(
A_{2}\right) ^{n_{2}}\left( A_{3}\right)
^{n_{3}}G_{T}(a_{1},a_{2},a_{3};n_{1},n_{2},n_{3})%
\begin{minipage}{4.5cm} \includegraphics[scale=.7]
{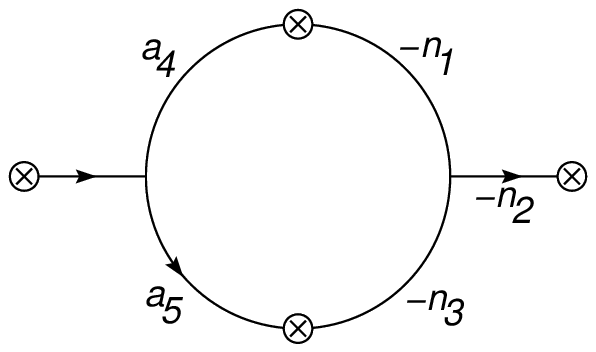} \end{minipage}
\end{equation}%
Using then the massless bubble one-loop functions $G_{A}$ \cite{IGo3}, we
quickly obtain the diagram MRE as:

\begin{equation}
\begin{array}{ll}
G= & \sum\limits_{n_{1},..,n_{3},\,l_{1}}\left( A_{1}\right) ^{n_{1}}\left(
A_{2}\right) ^{n_{2}}\left( A_{3}\right)
^{n_{3}}G_{T}(a_{1},a_{2},a_{3};n_{1},n_{2},n_{3})\times
G_{A}(a_{4}-n_{1},a_{5}-n_{3};n_{7}) \\
&  \\
&
\begin{array}{c}
\left( -n_{2}-n_{7}\right) \\
\begin{minipage}{2.9cm} \includegraphics[scale=.7]
{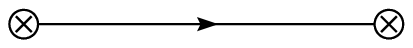} \end{minipage} \\
\end{array}%
\end{array}%
\end{equation}%
or equivalently:

\begin{equation}
G=\sum\limits_{n_{1},..,n_{3},\,n_{7}}\left( A_{1}\right) ^{n_{1}}\left(
A_{2}\right) ^{n_{2}}\left( A_{3}\right)
^{n_{3}}G_{T}(a_{1},a_{2},a_{3};n_{1},n_{2},n_{3})\times
G_{A}(a_{4}-n_{1},a_{5}-n_{3};n_{7})\;\frac{1}{\left( p^{2}\right)
^{-n_{2}-n_{7}}}.
\end{equation}%
The one-loop functions $G_{T}$ and $G_{A}$ are given by the expressions:

\begin{equation}
G_{T}(a_{1},a_{2},a_{3};n_{1},n_{2},n_{3})=\dfrac{(-1)^{-D/2}}{%
\prod\nolimits_{j=1}^{3}\Gamma (a_{j})}\sum\limits_{n_{4},..,n_{6}}\phi
_{n_{1},..,n_{6}}\;\dfrac{\prod\nolimits_{j=1}^{4}\Delta _{j}}{\Gamma (\frac{%
D}{2}+n_{1}+n_{2}+n_{3})},
\end{equation}%
whose constraints are:

\begin{equation}
\left\{
\begin{array}{l}
\Delta _{1}=\left\langle \frac{D}{2}+n_{1}+n_{2}+n_{3}+n_{4}+n_{5}+n_{6}%
\right\rangle , \\
\Delta _{2}=\left\langle a_{1}+n_{1}+n_{2}+n_{4}\right\rangle , \\
\Delta _{3}=\left\langle a_{2}+n_{2}+n_{3}+n_{5}\right\rangle , \\
\Delta _{4}=\left\langle a_{3}+n_{1}+n_{3}+n_{6}\right\rangle ,%
\end{array}%
\right.
\end{equation}%
and for $G_{A}$ we have:

\begin{equation}
G_{A}\left( a_{4}-n_{1},a_{5}-n_{3};n_{7}\right) =\dfrac{(-1)^{-D/2}}{\Gamma
(a_{4}-n_{1})\Gamma (a_{5}-n_{3})}\sum\limits_{n_{8},n_{9}}\phi
_{n_{7},..,n_{9}}\;\dfrac{\prod\nolimits_{j=5}^{7}\Delta _{j}}{\Gamma (\frac{%
D}{2}+n_{7})},
\end{equation}%
where the constraints become:

\begin{equation}
\left\{
\begin{array}{l}
\Delta _{5}=\left\langle \frac{D}{2}+n_{7}+n_{8}+n_{9}\right\rangle , \\
\Delta _{6}=\left\langle a_{4}-n_{1}+n_{7}+n_{8}\right\rangle , \\
\Delta _{7}=\left\langle a_{5}-n_{3}+n_{7}+n_{9}\right\rangle .%
\end{array}%
\right.
\end{equation}%
With this information we finally are able to write the diagram MRE as:

\begin{equation}
\fbox{$G=\dfrac{(-1)^{-D}}{\prod\nolimits_{j=1}^{3}\Gamma (a_{j})}%
\sum\limits_{n_{1},..,n_{9}}\phi _{n_{1},..,n_{9}}$\ $\dfrac{\left(
A_{1}\right) ^{n_{1}}\left( A_{2}\right) ^{n_{2}}\left( A_{3}\right)
^{n_{3}}\left( p^{2}\right) ^{n_{2}+n_{7}}}{\Gamma (\frac{D}{2}%
+n_{1}+n_{2}+n_{3})\Gamma (\frac{D}{2}+n_{7})}\dfrac{\prod%
\nolimits_{j=1}^{7}\Delta _{j}}{\Gamma (a_{4}-n_{1})\Gamma (a_{5}-n_{3})}$}
\label{t6}
\end{equation}%
This expansion contains $9\Sigma $ and $7\delta $, and therefore the
solution of this diagram corresponds to a double series and since we are
dealing with a propagator, the kinematical variable $\left( p^{2}\right) $
that is present does not appear explicitly as an argument in this series. We
expect that for this type of diagrams the solution will not be a single
term, since it does not correspond to an optimal topology (in the sense that
the multiplicity of the resulting series does not depend on the number of
loops or equivalently only depends on the invariants of the physical process
\cite{IGo2}).

In order to simplify the notation in $\left( \ref{t6}\right) $\thinspace\ we
take $A_{1}=A,$ $A_{2}=B$ and $A_{3}=C$. With these invariants six regions
of interest can be generated, each one with a different solution, but in the
limit $A=B=C=1$ they are all identical. These regions are given by:

\begin{equation}
\left\{
\begin{array}{l}
A<B<C, \\
A<C<B, \\
B<A<C, \\
B<C<A, \\
C<B<A, \\
C<A<B.%
\end{array}%
\right.
\end{equation}%
We arbitrarily take the region in which $\left( A<C<B\right) $. Then when we
extract the different terms of the MRE (bivalued hypergeometric functions)
we only sum those that have as one of their arguments the following simple
combinations:

\begin{equation*}
\left( \frac{A}{C}\right) ,\left( \frac{A}{B}\right) ,\left( \frac{C}{B}%
\right) ,
\end{equation*}%
or equivalently all the double combinations generated when combining these
simple combinations: $\left( \frac{A}{C},\frac{C}{B}\right) ,$ $\left( \frac{%
A}{B},\frac{C}{B}\right) ,$ etc. Given these conditions, the general
solution that we get starting from $\left( \ref{t6}\right) $\ is the
following:

\begin{equation}
G=g_{1}+g_{2}+g_{3}+g_{4}+g_{5}+g_{6}
\end{equation}%
where we have that:

\begin{equation}
\begin{array}{ll}
g_{1}= & \left( -1\right) ^{-D}\left( p^{2}\right) ^{D-a_{12345}}\dfrac{%
\Gamma \left( a_{13}-\frac{D}{2}\right) \Gamma \left( \frac{D}{2}%
-a_{23}\right) \Gamma \left( \frac{D}{2}-a_{1}\right) \Gamma \left(
a_{1345}-D\right) \Gamma \left( \frac{D}{2}-a_{5}\right) \Gamma \left(
D-a_{134}\right) }{\Gamma \left( a_{1}\right) \Gamma \left( a_{3}\right)
\Gamma \left( a_{5}\right) \Gamma \left( D-a_{123}\right) \Gamma \left(
a_{134}-\frac{D}{2}\right) \Gamma \left( \frac{3D}{2}-a_{1345}\right) } \\
&  \\
& \times \;A^{\frac{D}{2}-a_{13}}B^{-a_{2}} \\
&  \\
& \times \;F^{\substack{ 2:2:2  \\ 2:1:1}}\left(
\begin{array}{ccc}
\left\{ a_{2},\;\frac{D}{2}-a_{1}\right\} , & \left\{ 1-a_{5},\;\frac{D}{2}%
-a_{5}\right\} , &  \\
\left\{ \frac{3D}{2}-a_{1345},\;D+1-a_{1345}\right\} , & \left\{ 1-\frac{D}{2%
}+a_{23}\right\} , &
\end{array}%
\right. \\
&  \\
&
\begin{array}{lll}
&  &
\end{array}%
\left. \left.
\begin{array}{c}
\left\{ \frac{D}{2}+1-a_{134},\;D-a_{134}\right\} \\
\left\{ \frac{D}{2}+1-a_{13}\right\}%
\end{array}%
\right\vert
\begin{array}{c}
\dfrac{C}{B},\dfrac{A}{B}%
\end{array}%
\right) ,%
\end{array}%
\end{equation}

\bigskip

\begin{equation}
\begin{array}{ll}
g_{2}= & \left( -1\right) ^{-D}\left( p^{2}\right) ^{D-a_{12345}}\dfrac{%
\Gamma \left( \frac{D}{2}-a_{45}\right) \Gamma \left( D-a_{2345}\right)
\Gamma \left( a_{12345}-D\right) \Gamma \left( \frac{D}{2}-a_{13}\right)
\Gamma \left( a_{345}-\frac{D}{2}\right) }{\Gamma \left( a_{1}\right) \Gamma
\left( a_{2}\right) \Gamma \left( a_{3}\right) \Gamma \left( \frac{D}{2}%
\right) \Gamma \left( D-a_{123}\right) } \\
&  \\
& \times \;B^{D-a_{12345}}C^{a_{45}-\frac{D}{2}} \\
&  \\
& \times \;\overline{F}^{\substack{ 2:2:2  \\ 2:1:1}}\left( \left.
\begin{array}{ccc}
\left\{ D-a_{2345},\;\frac{D}{2}-a_{45}\right\} , & \left\{ 1-a_{4},\;\frac{D%
}{2}-a_{4}\right\} , & \left\{ a_{12345}-D,\;a_{345}-\frac{D}{2}\right\} \\
\left\{ \frac{D}{2}-a_{4},\;1-a_{4}\right\} , & \left\{ 1+a_{13}-\frac{D}{2}%
\right\} , & \left\{ \frac{D}{2}\right\}%
\end{array}%
\right\vert
\begin{array}{c}
\dfrac{A}{C},\dfrac{C}{B}%
\end{array}%
\right) ,%
\end{array}%
\end{equation}

\bigskip

\begin{equation}
\begin{array}{ll}
g_{3}= & \left( -1\right) ^{-D}\left( p^{2}\right) ^{D-a_{12345}}\dfrac{%
\Gamma \left( a_{13}-\frac{D}{2}\right) \Gamma \left( a_{12345}-D\right)
\Gamma \left( D-a_{1345}\right) \Gamma \left( \frac{3D}{2}%
-2a_{3}-a_{1245}\right) \Gamma \left( a_{345}-\frac{D}{2}\right) }{\Gamma
\left( a_{1}\right) \Gamma \left( a_{2}\right) \Gamma \left( a_{3}\right)
\Gamma \left( \frac{D}{2}\right) \Gamma \left( D-a_{123}\right) } \\
&  \\
& \times \;A^{\frac{D}{2}-a_{13}}B^{D-a_{12345}}C^{a_{1345}-\frac{D}{2}} \\
&  \\
& \times \;\overline{F}^{\substack{ 2:2:2  \\ 2:1:1}}\left(
\begin{array}{cc}
\left\{ D-a_{1345},\;\frac{3D}{2}-2a_{3}-a_{1245}\right\} , & \left\{ \frac{D%
}{2}+1-a_{134},\;D-a_{134}\right\} , \\
\left\{ D-a_{134},\;\frac{D}{2}+1-a_{134}\right\} , & \left\{ \frac{D}{2}%
+1-a_{13}\right\} ,%
\end{array}%
\right. \\
&  \\
&
\begin{array}{lll}
&  &
\end{array}%
\left. \left.
\begin{array}{c}
\left\{ a_{12345}-D,\;a_{345}-\frac{D}{2}\right\} \\
\left\{ \frac{D}{2}\right\}%
\end{array}%
\right\vert
\begin{array}{c}
\dfrac{A}{C},\dfrac{C}{B}%
\end{array}%
\right) ,%
\end{array}%
\end{equation}

\bigskip

\begin{equation}
\begin{array}{ll}
g_{4}= & \left( -1\right) ^{-D}\left( p^{2}\right) ^{D-a_{12345}}\dfrac{%
\Gamma \left( \frac{D}{2}-a_{13}\right) \Gamma \left( \frac{D}{2}%
-a_{23}\right) \Gamma \left( a_{123}-\frac{D}{2}\right) \Gamma \left( a_{45}-%
\frac{D}{2}\right) \Gamma \left( \frac{D}{2}-a_{5}\right) \Gamma \left(
\frac{D}{2}-a_{4}\right) }{\Gamma \left( a_{1}\right) \Gamma \left(
a_{2}\right) \Gamma \left( a_{5}\right) \Gamma \left( D-a_{45}\right) \Gamma
\left( D-a_{123}\right) } \\
&  \\
& \times \;A^{\frac{D}{2}-a_{13}}B^{D-a_{12345}}C^{a_{1345}-\frac{D}{2}} \\
&  \\
& \times \;F^{\substack{ 2:2:2  \\ 2:1:1}}\left( \left.
\begin{array}{ccc}
\left\{ a_{3},\;a_{123}-\frac{D}{2}\right\} , & \left\{ 1-a_{4},\;\frac{D}{2}%
-a_{4}\right\} , & \left\{ 1-a_{5},\;\frac{D}{2}-a_{5}\right\} \\
\left\{ D-a_{45},\;1+\frac{D}{2}-a_{45}\right\} , & \left\{ 1-\frac{D}{2}%
+a_{13}\right\} , & \left\{ 1-\frac{D}{2}+a_{23}\right\}%
\end{array}%
\right\vert
\begin{array}{c}
\dfrac{A}{B},\dfrac{C}{B}%
\end{array}%
\right) ,%
\end{array}%
\end{equation}

\bigskip

\begin{equation}
\begin{array}{ll}
g_{5}= & \left( -1\right) ^{-D}\left( p^{2}\right) ^{D-a_{12345}}\dfrac{%
\Gamma \left( a_{13}-\frac{D}{2}\right) \Gamma \left( a_{23}-\frac{D}{2}%
\right) \Gamma \left( \frac{D}{2}-a_{3}\right) \Gamma \left(
D-a_{134}\right) \Gamma \left( D-a_{235}\right) }{\Gamma \left( a_{1}\right)
\Gamma \left( a_{2}\right) \Gamma \left( a_{3}\right) \Gamma \left(
2D-a_{1245}-2a_{3}\right) \Gamma \left( a_{134}-\frac{D}{2}\right) } \\
&  \\
& \times \dfrac{\Gamma \left( a_{1245}+2a_{3}-\frac{3D}{2}\right) }{\Gamma
\left( a_{235}-\frac{D}{2}\right) }\;A^{\frac{D}{2}-a_{13}}B^{a_{3}-\frac{D}{%
2}}C^{\frac{D}{2}-a_{23}} \\
&  \\
& \times \;F^{\substack{ 2:2:2  \\ 2:1:1}}\left(
\begin{array}{cc}
\left\{ D-a_{123},\;\frac{D}{2}-a_{3}\right\} , & \left\{ \frac{D}{2}%
+1-a_{235},\;D-a_{235}\right\} , \\
\left\{ 2D-a_{1245}-2a_{3},\;\frac{3D}{2}+1-2a_{3}-a_{1245}\right\} , &
\left\{ \frac{D}{2}+1-a_{23}\right\} ,%
\end{array}%
\right. \\
&  \\
&
\begin{array}{lll}
&  &
\end{array}%
\left. \left.
\begin{array}{c}
\left\{ \frac{D}{2}+1-a_{134},\;D-a_{134}\right\} \\
\left\{ \frac{D}{2}+1-a_{13}\right\}%
\end{array}%
\right\vert
\begin{array}{c}
\dfrac{C}{B},\dfrac{A}{B}%
\end{array}%
\right) ,%
\end{array}%
\end{equation}%
and finally:

\begin{equation}
\begin{array}{ll}
g_{6}= & \left( -1\right) ^{-D}\dfrac{\Gamma \left( \frac{D}{2}%
-a_{13}\right) \Gamma \left( a_{23}-\frac{D}{2}\right) \Gamma \left( \frac{D%
}{2}-a_{2}\right) \Gamma \left( \frac{D}{2}-a_{4}\right) \Gamma \left(
a_{2345}-D\right) \Gamma \left( D-a_{235}\right) }{\Gamma \left(
a_{2}\right) \Gamma \left( a_{3}\right) \Gamma \left( a_{4}\right) \Gamma
\left( D-a_{123}\right) \Gamma \left( a_{235}-\frac{D}{2}\right) \Gamma
\left( \frac{3D}{2}-a_{2345}\right) } \\
&  \\
& \times \;B^{-a_{1}}C^{\frac{D}{2}-a_{23}} \\
&  \\
& \times \;F^{\substack{ 2:2:2  \\ 2:1:1}}\left(
\begin{array}{cc}
\left\{ a_{1},\;\frac{D}{2}-a_{2}\right\} , & \left\{ 1-a_{4},\;\frac{D}{2}%
-a_{4}\right\} , \\
\left\{ \frac{3D}{2}-a_{2345},\;1+D-a_{2345}\right\} , & \left\{ 1-\frac{D}{2%
}+a_{13}\right\} ,%
\end{array}%
\right. \\
&  \\
&
\begin{array}{lll}
&  &
\end{array}%
\left. \left.
\begin{array}{c}
\left\{ 1+\frac{D}{2}-a_{235},\;D-a_{235}\right\} \\
\left\{ 1+\frac{D}{2}-a_{23}\right\}%
\end{array}%
\right\vert
\begin{array}{c}
\dfrac{A}{B},\dfrac{C}{B}%
\end{array}%
\right) .%
\end{array}%
\end{equation}%
The previous results have been written in terms of the functions $F$ and $%
\overline{F}$ (for more information see Ref. \cite{IGo2}\ ), which are
defined as:
\begin{equation}
\overline{F}^{\substack{ p:r:u  \\ q:s:v}}\left( \left.
\begin{array}{ccc}
\left\{ \gamma _{1},..,\gamma _{p}\right\} & \left\{ a_{1},..,a_{r}\right\}
& \left\{ c_{1},..,c_{u}\right\} \\
\left\{ \beta _{1},..,\beta _{q}\right\} & \left\{ b_{1},..,b_{s}\right\} &
\left\{ d_{1},..,d_{v}\right\}%
\end{array}%
\right\vert x,y\right) =\sum\limits_{n,m}^{\infty }\dfrac{%
\prod\limits_{j=1}^{p}(\gamma
_{j})_{n-m}\prod\limits_{j=1}^{r}(a_{j})_{n}\prod\limits_{j=1}^{u}(c_{j})_{m}%
}{\prod\limits_{j=1}^{q}(\beta
_{j})_{n-m}\prod\limits_{j=1}^{s}(b_{j})_{n}\prod\limits_{j=1}^{v}(d_{j})_{m}%
}\dfrac{x^{n}}{n!}\dfrac{y^{m}}{m!},
\end{equation}%
and similarly we also have:

\begin{equation}
F^{\substack{ p:r:u  \\ q:s:v}}\left( \left.
\begin{array}{ccc}
\left\{ \gamma _{1},..,\gamma _{p}\right\} & \left\{ a_{1},..,a_{r}\right\}
& \left\{ c_{1},..,c_{u}\right\} \\
\left\{ \beta _{1},..,\beta _{q}\right\} & \left\{ b_{1},..,b_{s}\right\} &
\left\{ d_{1},..,d_{v}\right\}%
\end{array}%
\right\vert x,y\right) =\sum\limits_{n,m}^{\infty }\dfrac{%
\prod\limits_{j=1}^{p}(\gamma
_{j})_{n+m}\prod\limits_{j=1}^{r}(a_{j})_{n}\prod\limits_{j=1}^{u}(c_{j})_{m}%
}{\prod\limits_{j=1}^{q}(\beta
_{j})_{n+m}\prod\limits_{j=1}^{s}(b_{j})_{n}\prod\limits_{j=1}^{v}(d_{j})_{m}%
}\dfrac{x^{n}}{n!}\dfrac{y^{m}}{m!}.
\end{equation}%
This last series is called Kamp\'{e} de F\'{e}riet function. Finally the
solution is obtained by simply taking $A=B=C=1$.

We have solved this diagram in a modular or loop by loop form. In order to
appreciate the advantages of this way of applying IBFE, it is important to
compare the MRE $\left( \ref{t6}\right) $\ with the MRE of the same diagram
obtained by the application of IBFE to the parametric integral of the
complete diagram. For this purpose let us consider first Schwinger's
parametric representation of the complete diagram, which is given by:

\begin{equation}
G=\dfrac{(-1)^{-D}}{\prod\limits_{j=1}^{5}\Gamma (a_{j})}\int\limits_{0}^{%
\infty }d\overrightarrow{x}\;\frac{\exp \left[ -\dfrac{F}{\left(
x_{1}+x_{2}\right) \left( x_{3}+x_{4}+x_{5}\right) +x_{3}\left(
x_{4}+x_{5}\right) }\right] }{\left[ \left( x_{1}+x_{2}\right) \left(
x_{3}+x_{4}+x_{5}\right) +x_{3}\left( x_{4}+x_{5}\right) \right] ^{\frac{D}{2%
}}},  \label{t11}
\end{equation}%
where the polynomial $F$ is given by:

\begin{equation}
F=\left[ x_{1}x_{2}\left( x_{3}+x_{4}+x_{5}\right) +x_{4}x_{5}\left(
x_{1}+x_{2}\right) +x_{3}x_{4}x_{5}+x_{1}x_{3}x_{5}+x_{2}x_{3}x_{4}\right]
\;p^{2}.
\end{equation}%
If we now obtain the diagram MRE using $\left( \ref{t11}\right) $\thinspace
we can compare (\textit{Table\ I}) this MRE with MRE $\left( \ref{t6}\right)
$:

\begin{equation}
\begin{tabular}{lll}
\hline
& IBFE complete diagram & IBFE modular \\ \hline
Multiplicity multiregion series $\left( {\sigma }\right) $ &
\multicolumn{1}{c}{12} & \multicolumn{1}{c}{9} \\
Kronecker deltas of the expansion $\left( {\delta }\right) $ &
\multicolumn{1}{c}{8} & \multicolumn{1}{c}{7} \\
Multiplicity resulting series $\left( {\sigma -\delta }\right) $ &
\multicolumn{1}{c}{4} & \multicolumn{1}{c}{2} \\
Possible contributions to the solutions $\left( {C}_{\delta }^{\sigma
}\right) $ & \multicolumn{1}{c}{495} & \multicolumn{1}{c}{36} \\ \hline
\end{tabular}
\tag{$Table\;I$}
\end{equation}%
We find that the MRE obtained in terms of the one-loop functions $G_{T}$ is
considerably reduced with respect to the one that we get when IBFE is
applied to both loops of the diagram simultaneously, not only in the number
of sums and Kronecker deltas, but also the difference between them is
reduced, which implies that the multiplicity of the resulting series is
less, a nontrivial fact that simplifies considerably the complexity of the
solutions.

\section{Commentaries}

\qquad The calculation technology that we have presented here can be useful
in order to find minimal expressions for the MRE of diagrams that contain
triangular subgraphs, such as those that we see in the following figure:

\begin{equation*}
\begin{minipage}{11.5cm} \includegraphics[scale=.8]
{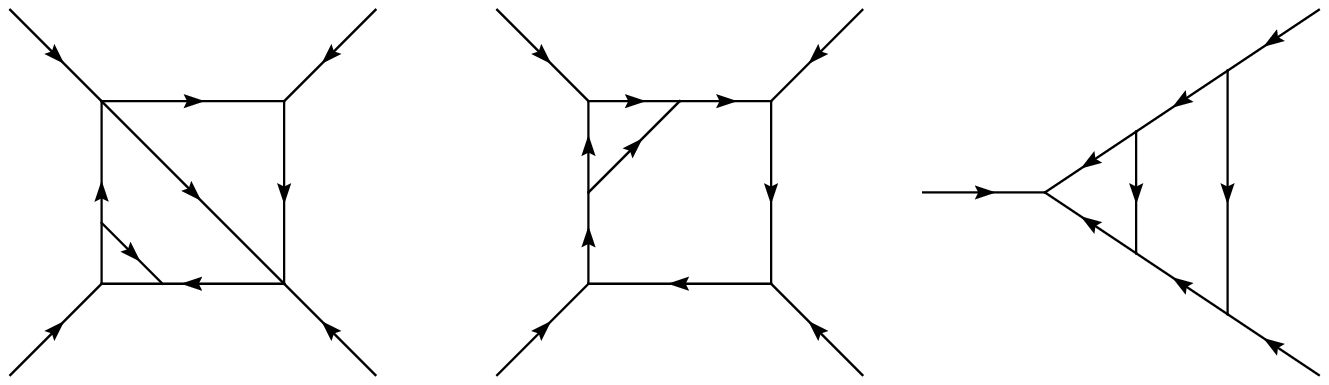} \end{minipage}\;\text{etc.}
\end{equation*}%
The loop function $G_{T}$ is very useful since it opens up the possibility
of evaluating other families of graphs and to obtain solutions with less
complexity and less number of terms when compared with the application of
IBFE to a parametric integral of the complete diagram. Although we have
presented here the one-loop function $G_{T}$ for the massless case, it is
certainly possible to consider also massive propagators and define new loop
functions for these cases. It will be always possible to use here the same
reduction procedure for diagrams that contain bubbles or triangles that we
have used for the massless cases.

The same idea can be extended to consider one-loop functions for modules
that contain four or more propagators. Nevertheless, the reductions are no
longer so simple as for the bubble and triangle subgraphs. The important
point is that the modular IBFE application, particularly when it is done
loop by loop, allows to obtain a minimal representation for the diagram MRE,
from which the analytical solution follows.

\section{Conclusions}

\qquad The general solutions that can be obtained in the evaluation of
Feynman diagrams always correspond to multivariable hypergeometric series,
whose multiplicity can be deduced directly from the MRE that represents the
diagram. One of the characteristics of the IBFE technique is that it has a
lower bound that fixes the multiplicity $\mu $ of these hypergeometric
series, and which is related to the minimal number of invariants $\left(
N\right) $ that characterizes a specific Feynman diagram, described through
the formula:

\begin{equation}
\mu \geqslant \left( N-1\right) .
\end{equation}%
The greater-equal sign is here included since besides the number of
invariants one must consider also the topological family to which the
diagram belongs. If this is not the optimal for IBFE application, in the
sense described in \cite{IGo2}, then the multiplicity is also increased
according to the number of loops in the diagram. This multiplicity can be
directly recognized from the MRE of the diagram, and it is equal to the
difference between the number of sums and constraints (Kronecker deltas)
that are present. Solutions with 0 multiplicity (which in reality is not a
series, and contains only one term), or multiplicity 1 or 2, correspond to
series whose properties are extensively discussed in the literature \cite%
{WBa, LSl, LGr, HEx}. Nevertheless, for triple or higher multiplicity
series, the information is rather limited and it is therefore important to
have methods that allow to reduce the diagram MRE, not only in the sense of
lowering the number of sums and deltas, but also to reduce their difference,
which is something that we have shown here can be done modularly (see
Example II).

The idea of the modular application of IBFE can be easily extended to
one-loop modules or subgraphs that contain four or more propagators. The
result will be in general a reduced MRE, that will also contain simpler
solution, i.e. hypergeometric functions of lower multiplicity. In the same
way, the idea of using fictitious invariants for selecting the correct
solution associated to a diagram can be also generalized to all the cases in
which hypergeometric functions with unit arguments are generated.

The modular application of the integration technique IBFE has shown to be a
powerful calculational tool, not only for its simplicity but also because it
allows to calculate many families of graphs. Here we have seen that the
systematical graphical procedure that we have implemented for triangular
modules, makes the reduction just as simple as in the case of bubbles
contained in a diagram.

\subparagraph{\protect\bigskip}

\textbf{Acknowledgements :}

We acknowledge support from Fondecyt under Grant No. 3080029.

\end{document}